\documentclass{aa}
\include{epsf}
\usepackage{graphics}

\begin{document}

\thesaurus{02(03.13.2; 12.07.1)}

\title{Weak weak lensing: correcting weak shear measurements\\
accurately for PSF anisotropy} 
\author{Konrad Kuijken\thanks{Visiting scientist, Dept. of Theoretical
Physics, Univ. of the Basque Country}}
\institute{Kapteyn Astronomical Institute} 
\date{Received ??? / Accepted ???}
\titlerunning{Weak weak lensing}
\maketitle

\begin{abstract}
We have developed a new technique for weak lensing analysis, with
which the effect of the point spread function (PSF) on small galaxy
images can
be corrected for accurately. Rather than relying on weighted second
moments of detected images, which we show can leave residuals at the
level of a percent in the shear, we directly fit (stacked or individual)
galaxy images
as PSF-convolved, sheared circular sources. We show by means of
simulations that this technique is able to recover shears well below
the percent level for a variety of PSF shapes, and that its noise
properties are similar to existing methods. 
\end{abstract}

\section{Introduction}

Gravitational lensing is one of the most powerful and direct methods
for studying the gravitational potentials of massive objects in the
universe. An important type of lensing study is `weak lensing.' It is
the study of mild systematic distortions of background sources as
their light rays are perturbed by gravitational fields on their way to
us. Weak lensing has already provided important results in the study
of galaxy clusters (e.g., Tyson et al.~\cite{tvw}; Bonnet et
al.~\cite{q2345}; Fahlman et al.~\cite{fahlman}; Squires et
al.~\cite{a2163}; Fischer et al.~\cite{q0957}; Clowe et
al.~\cite{clowe}; Hoekstra et al.~\cite{hoekstra}), halos of
individual galaxies (e.g., Brainerd et al.~\cite{brainerd}), and
large-scale structure (Schneider et al.~\cite{schneider98}). As the
techniques are becoming better understood, research is progressing to
the search for weaker and weaker distortions, which would enable the outer
regions of galaxy clusters and galaxy halos, as well as lensing
signals from large-scale structure (e.g., Jain \&
Seljak~\cite{jain97}; Kaiser~\cite{kaiser98}), to be studied.

To be able to detect such very weak signals, it is important to
accurately remove the dominant systematic effect affecting weak
lensing measurements: anisotropy of, and smearing by, the point-spread
function (PSF). In this paper we will first investigate the limits of
the most commonly-used technique for weak lensing analysis, devised by
Kaiser et al.~(\cite{ksb}, henceforth KSB). We will show that after
PSF anisotropy correction, residual effects on the order of 1\% shear
are difficult to avoid with this method, even for moderately elongated
PSF's. Since the ability to detect percent signals is important for a
variety of scientific questions, we have therefore devised a new
method which does not have such residuals, but which nevertheless has
noise properties comparable to those of the KSB method.

There are several other methods for PSF anisotropy correction in the
literature. The Autocorrelation Function method of Van Waerbeke et
al.~(\cite{waerbeke}) is a variant of the KSB method in which not
individual galaxy images, but the autocorrelation function of many of
them, is analyzed. The Bonnet \& Mellier~(\cite{bm}) method uses a
different aperture weighting function from KSB, and treats the PSF
convolution as a shear term. Fisher \& Tyson~(\cite{ft97}) convolve
the image with a kernel constructed to make the PSF rounder again.
A more sophisticated such kernel has recently been presented 
by Kaiser~(\cite{k99}).

\section{Limitations of the KSB formalism for PSF anisotropy correction}

We first examine the Kaiser et al.~(\cite{ksb}) method, following our
earlier limited investigation in the context of analysis of Hubble
Space Telescope images (Hoekstra et al.~\cite{hoekstra}, Appendix~D). 

\subsection{The KSB method}

The technique of weak (or statistical) lensing involves measuring the
systematic, gravitationally induced, distortion of background images
behind a gravitational lens. In the weak lensing regime small
background images are distorted by a shear $(\gamma_1,\gamma_2)$ and a
convergence $\kappa$, whose combined effect is represented by the
mapping
\begin{eqnarray}
\pmatrix{x\cr y}
&\to
\pmatrix{1-\kappa-\gamma_1 & -\gamma_2 \cr -\gamma_2 & 1-\kappa+\gamma_1}
\pmatrix{x\cr y}\cr
&\equiv
(1-\kappa)
\pmatrix{1-g_1 & -g_2 \cr -g_2 & 1+g_1}
\pmatrix{x\cr y}
\label{eq:dist}
\end{eqnarray}
where $g_i=\gamma_i/(1-\kappa)$.  For simplicity, in what follows we
neglect $\kappa$ as it is small in the weak lensing regime, and
pretend we are deriving the true shear $\gamma$ instead of the reduced
shear $g$. Thus, our results on shape measurements are valid, but their
interpretation as a lensing signal may require consideration of the
$(1-\kappa)$ factor. Our analysis makes no assumptions on the
smallness of $g_i$, though.

Kaiser, Squires and Broadhurst (1995, henceforth KSB) describe a
method for recovering $(\gamma_1,\gamma_2)$ from images of distant
galaxies. Essentially, they derive galaxy ellipticities from weighted
second moments of the observed images, and then correct these for the
effects of the weight function and of smearing by the point spread
function (PSF). By averaging over many galaxies, which are assumed to
be intrinsically randomly oriented, the effect of individual galaxy
ellipticities should average out, leaving the systematic lensing
signal. The KSB method has proved to be very effective, especially in
the study of galaxy cluster potentials.

\begin{figure}
\resizebox{\hsize}{!}{\includegraphics{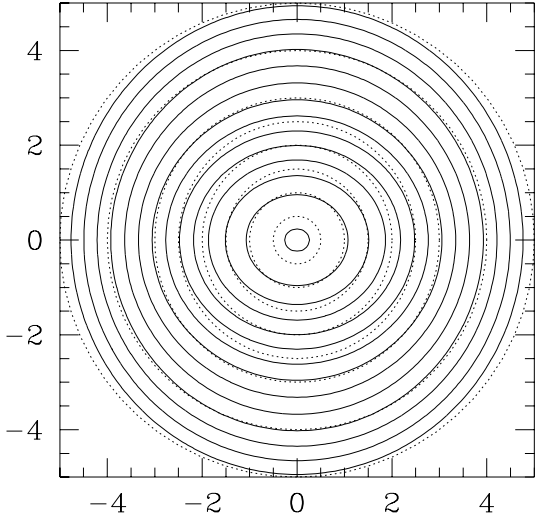}}
\caption{An example of a PSF with radially varying ellipticity, but
zero overall polarization. (Eq.~\ref{eq:psfone} with $\delta=0.3$).
Contours differ by a factor of $2^{1/2}$. The dotted curves are
circles, shown for comparison.}
\label{fig:modelpsf}
\end{figure}

\begin{figure}
\resizebox{\hsize}{!}{\includegraphics{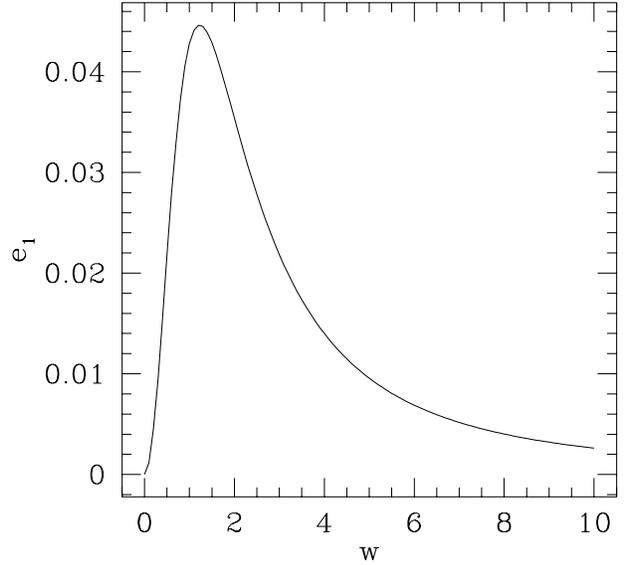}}
\caption{The polarization of the PSF of Figure~\ref{fig:modelpsf} as a
function of the weight function's Gaussian radius $w$.  The plot shows
that with compact weight functions such as those that are required to
suppress photon noise, even mildly elongated PSF's may erroneously
yield a polarization of several percent.}
\label{fig:dcoeff}
\end{figure}

KSB define various ``polarizabilities'', which express the ratio
between an input distortion (gravitational shear or PSF anisotropy)
and the measured polarization
\begin{equation}
e=\left({I^w_{xx}-I^w_{yy}\over I^w_{xx}+I^w_{yy}},{2I^w_{xy}\over
I^w_{xx}+I^w_{yy}}\right)
\end{equation}
of an image $f(x,y)$. These polarizations are derived from weighted second
moments 
\begin{equation}
\label{eq:wtfn}
I^w_{xx}=\int f(x,y) x^2 W(r) dx dy,\qquad\hbox{etc.,}
\end{equation}
 of the image intensities, where $W$ is a weight function which goes
to zero at large radii. The weight function is required as otherwise
the sky noise in the outer parts of the image dominates the measured
moment. The significance of the measurement is optimized by taking the
weight function to be relatively compact, of a size comparable to the
image itself.

Details of the method can be found in KSB, and in Hoekstra et al.\
(1998, henceforth HFKS), where a few small errors in the formulae of
KSB were corrected. For the purposes of the present paper, it is
sufficient to know that in the KSB formalism, the ``smear
polarizability" $P^{\rm sm}$ defines the ratio between the PSF
anisotropy $p=(I_{xx}-I_{yy},2I_{xy})$, constructed from the
unweighted second moments $I_{ij}$ of the (normalized) PSF, and the
resulting change in image polarization $e$. The ``shear
polarizability" $P^{\rm sh}$ is the ratio between the applied shear
$(\gamma_1,\gamma_2)$ and the resulting change in the image
polarization. KSB show how the polarizabilities can be derived from
higher weighted moments of the observed PSF and galaxy images.

\subsection{How accurate is KSB?}

In the context of ground-based cluster weak lensing, the KSB method
works well. Nevertheless, it does involve some approximations. Now
that weaker and weaker signals are of interest, it is therefore
important to understand the limitations of the method.  As already
discussed by HFKS, for strongly non-Gaussian PSF's the KSB method does
not completely correct PSF anisotropy. This is particularly true when
analyzing small galaxies in deep HST images, where it turns out that
the choice of weight function in eq.~\ref{eq:wtfn} is important.

A simple PSF model can be used to illustrate why such residuals are,
at some level, unavoidable.  Consider the following PSF:
\begin{equation}
\label{eq:psfone}
P(x,y)=G(1+\delta,1)+G(4-\delta,4),
\end{equation}
where $G(a^2,b^2)$ is a unit-area Gaussian of $x$- and $y$-dispersions
$a$ and $b$. $\delta$ is a small parameter. The case $\delta=0.3$ is
plotted in Figure~\ref{fig:modelpsf}.

The PSF of equation \ref{eq:psfone} has exactly zero anisotropy $p$:
the second moments in $x$ and in $y$ are equal. However, the
ellipticity of the PSF varies with radius, which means that the
{\em weighted} second moments are not equal: weighting the central parts
more will enhance the $x$-moment preferentially.  In fact, it is easy
to show that the polarization constructed with weighted moments is
$O(\delta)$. The precise result for a Gaussian weight function
$W=\exp(-{1\over2}r^2/w^2)$ is
\begin{equation}
e_1=  {\frac{9\,{w^2}\,\left( 7 + 5\,{w^2} + {w^4} \right) \,\delta}
     {2\,\left( 1+{w^2}\right) \,
       \left( 4 + {w^2} \right) \,
       \left( 20 + 16\,{w^2} + 5\,{w^4} \right) }} +
     {{{\rm O}(\delta^2)}}.
\label{eq:dcoeff}
\end{equation}
The polarization of the $\delta=0.3$ PSF plotted in
Figure~\ref{fig:modelpsf} is plotted in Figure~\ref{fig:dcoeff}. It
has a value of 0.03 near $w=2$, roughly the radius of maximum
significance which should be used to minimize photon noise in the
polarization measurement. 

The KSB polarizabilities are derived assuming that the PSF can be
written as the convolution of a {\em compact} anisotropic part with an
{\em extended} circular part (KSB eq.~A1). This assumption allows the
anisotropy to be characterized in terms of $p$ only. However, our
example shows that this assumption may be too restrictive: it
effectively couples the radial intensity profile of the PSF with its
ellipticity profile. For example, a single Gaussian with constant
ellipticity can be written as such a convolution, but a sum of two
elliptical Gaussian such as the PSF of eq.~\ref{eq:psfone} cannot. The
systematic errors that arise are the result of this.

\section{A new method}

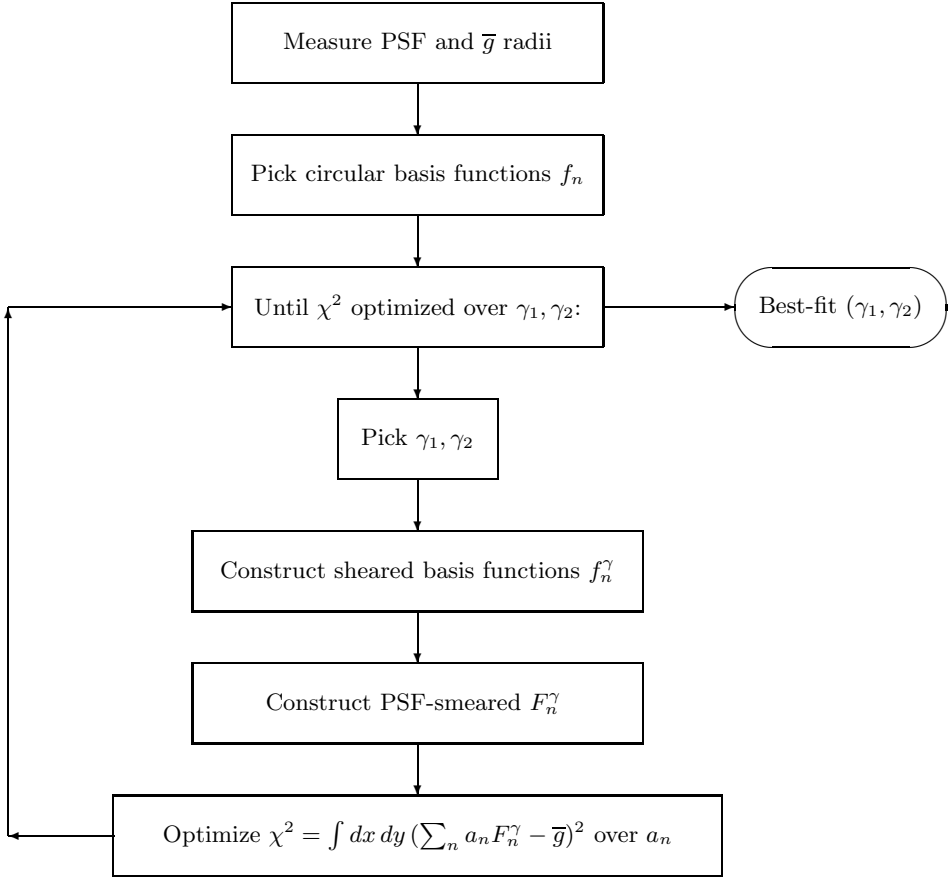
\begin{figure*}
\begin{picture}(250,350)(0,20)
\put(160,330){\framebox(140,30){Measure PSF and ${\overline{g}}$ radii}}
\put(230,330){\vector(0,-1){20}}
\put(160,280){\framebox(140,30){Pick circular basis functions $f_n$}}
\put(230,280){\vector(0,-1){20}}
\put(160,230){\framebox(140,30){Until $\chi^2$ optimized over $\gamma_1,\gamma_2$:}}
\put(230,230){\vector(0,-1){20}}
\put(200,180){\framebox(60,30){Pick $\gamma_1,\gamma_2$}}
\put(230,180){\vector(0,-1){20}}
\put(145,130){\framebox(170,30){Construct sheared basis functions $f_n^\gamma$}}
\put(230,130){\vector(0,-1){20}}
\put(145,080){\framebox(170,30){Construct PSF-smeared $F_n^\gamma$ }}
\put(230,080){\vector(0,-1){20}}
\put(115,030){\framebox(230,30){Optimize $\chi^2=\int dx\, dy\,(\sum_n a_n F_n^\gamma-{\overline{g}})^2$
over $a_n$}}
\put(115,045){\vector(-1,0){40}}
\put(075,045){\vector(0,1){200}}
\put(075,245){\vector(1,0){85}}
\put(390,245){\oval(80,30)}
\put(300,245){\vector(1,0){50}}
\put(350,230){\makebox(80,30){Best-fit $(\gamma_1,\gamma_2)$}}
\end{picture}
\caption{The schematic algorithm used to derive the shear from an
observed mean galaxy and PSF image.}
\label{fig:algo}
\end{figure*}

Here we present a new method, with which the PSF effects can be
corrected for with greater accuracy. The essence of the method is not
to work with the moments of the observed images; instead each image is
fit directly as a PSF-convolved, sheared circular source of unknown
radial profile.

Assume for the moment that we have managed to sum the images of many
galaxies into an `average galaxy' image ${\overline{g}}(x,y)$.
Analysing a stacked galaxy image is similar to the approach discussed
by Lombardi \& Bertin (\cite{lb}), who average image second moments
before corrections are applied. It differs from methods such as KSB or
Bonnet \& Mellier (\cite{bm}) in which galaxies are individually
corrected for PSF effects before they are combined to produce a shear
estimate. 

Intrinsically, ${\overline{g}}$ is circular if the galaxies are
randomly oriented, but the image we observe has been distorted first
by gravitational lensing shear, then by the atmospheric seeing, and
finally by the camera optics.  The observed ${\overline{g}}$ is
therefore a sheared circular source, convolved with a (known) PSF. We
therefore fit ${\overline{g}}$ directly to such a model, with the
minimum of further assumptions. This approach addresses the apparent
difficulty in the KSB methodology in the case of radially changing
ellipticity profiles: a sheared circular source has constant
ellipticity at all radii, and so after convolution with the PSF only a
subset of ``allowed'' ellipticity profiles remain.  

Assuming that the PSF is known, e.g., from analysis of star images in
the field, the model for ${\overline{g}}$ is specified by an unknown
radial brightness profile, and by the shear parameters
$(\gamma_1,\gamma_2)$ that we are interested in. In practice we model
the radial profile as the superposition of several Gaussians of
different fixed widths, and unknown amplitude. We have found that the
following recipe for assigning the basis functions gives good results:
(i) determine the best-fit circular Gaussian radii to the observed PSF
and galaxy images, $r_{\rm PSF}$ and $r_{\rm GAL}$. (ii) Take
$r=(r_{\rm GAL}^2-r_{\rm PSF}^2)^{1/2}$ as an estimate for the
intrinsic radius of ${\overline{g}}$.  (iii) Use four components to
describe the radial profile of $\overline{g}$, with
Gaussian radii $(0.5,1,2,4)\times r$.

The algorithm is laid out in Figure~\ref{fig:algo}. We now describe
the results of tests to verify the accuracy of the PSF anisotropy
correction, and to investigate how well it fares in the presence of
noise in the images. 

\subsection{Simulations in the absence of noise}

We tested how well PSF anisotropy can be corrected for by considering
the case where there is no gravitational shear, only a range of PSF
shapes of varying anisotropy. An accurate analysis should yield zero
shear after correction for the PSF. On a large number of model images,
described below, we compared the results of the algorithm of
Figure~\ref{fig:algo} with those from the KSB algorithm as described
in HFKS (implying in particular that the same weight function is used
in the derivation of polarizations and polarizabilities of galaxy and
PSF images). The weight function was taken to be the best-fit circular
Gaussian to the post-seeing galaxy image.

The algorithm described in this paper directly yields an estimate for
the shear. In the comparisons, the KSB galaxy polarization after
seeing anisotropy correction was divided by the ``pre-seeing shear
polarizability" $P^\gamma$, for which we use the expression given by
Luppino \& Kaiser (1997). 

\subsubsection{Double-Gaussian images and PSF}
\begin{figure}
\resizebox{\hsize}{!}{\includegraphics{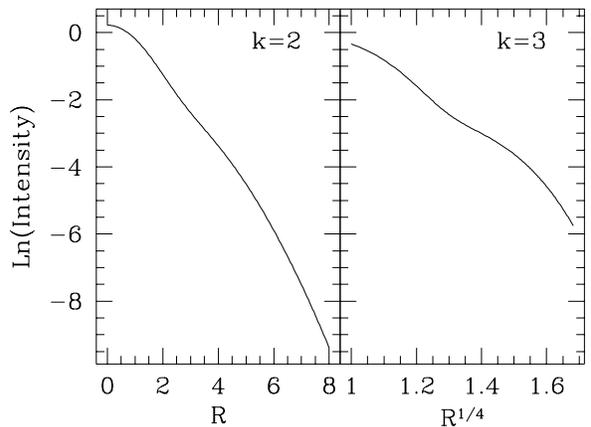}}
\caption{The double-Gaussian profiles used in the modeling in this
paper. Left, the $k=2$ profile is plotted logarithmically to show its
similarity to an exponential profile; right the $k=3$ profile is
plotted logarithmically vs. $r^{1/4}$ to show it is similar to a de
Vaucouleurs model.}
\label{fig:kprofiles}
\end{figure}

In most of our simulations, we modeled the round average galaxy images as
\begin{equation}
\label{eq:gal}
{\overline{g}}=G(r_g)+G(k_g r_g)
\end{equation}
where $G(\sigma)$ is a unit-integral Gaussian of dispersion $\sigma$
(a double-Gaussian PSF was also considered by ???REF???). The
parameter $k_g$ is unity for a Gaussian profile, and is larger for
more radially extended profiles. A reasonable, though admittedly crude,
approximation to an exponential profile is given by setting $k_g=2$,
while $k_g=3$ gives a reasonable approximation to a de vaucouleurs
profile (Figure \ref{fig:kprofiles}).

The PSF's were modeled in a similar way, but with anisotropy. Writing
now $G(a,b)$ for a Gaussian with $x$- and $y$-dispersions $a$ and $b$,
we have
\begin{equation}
\label{eq:psf}
PSF=
G\left(r_p,(1-\epsilon_1)r_p\right)+G\left(k_pr_p,(1-\epsilon_2)k_pr_p\right).
\end{equation}
Again we introduced a shape parameter $k_p$, but we also included
ellipticities $\epsilon_i$ for the two components. We considered three
kinds of PSF ellipticity profile: we either set
$\epsilon_1=\epsilon_2$ (constant ellipticity with radius), or set one
of the $\epsilon$'s to zero, to give a radial increase or decrease of
the PSF anisotropy. These three possibilities, though by no means
exhaustive, form a representative set of PSF's.


\begin{figure*}
\resizebox{\hsize}{!}{\includegraphics{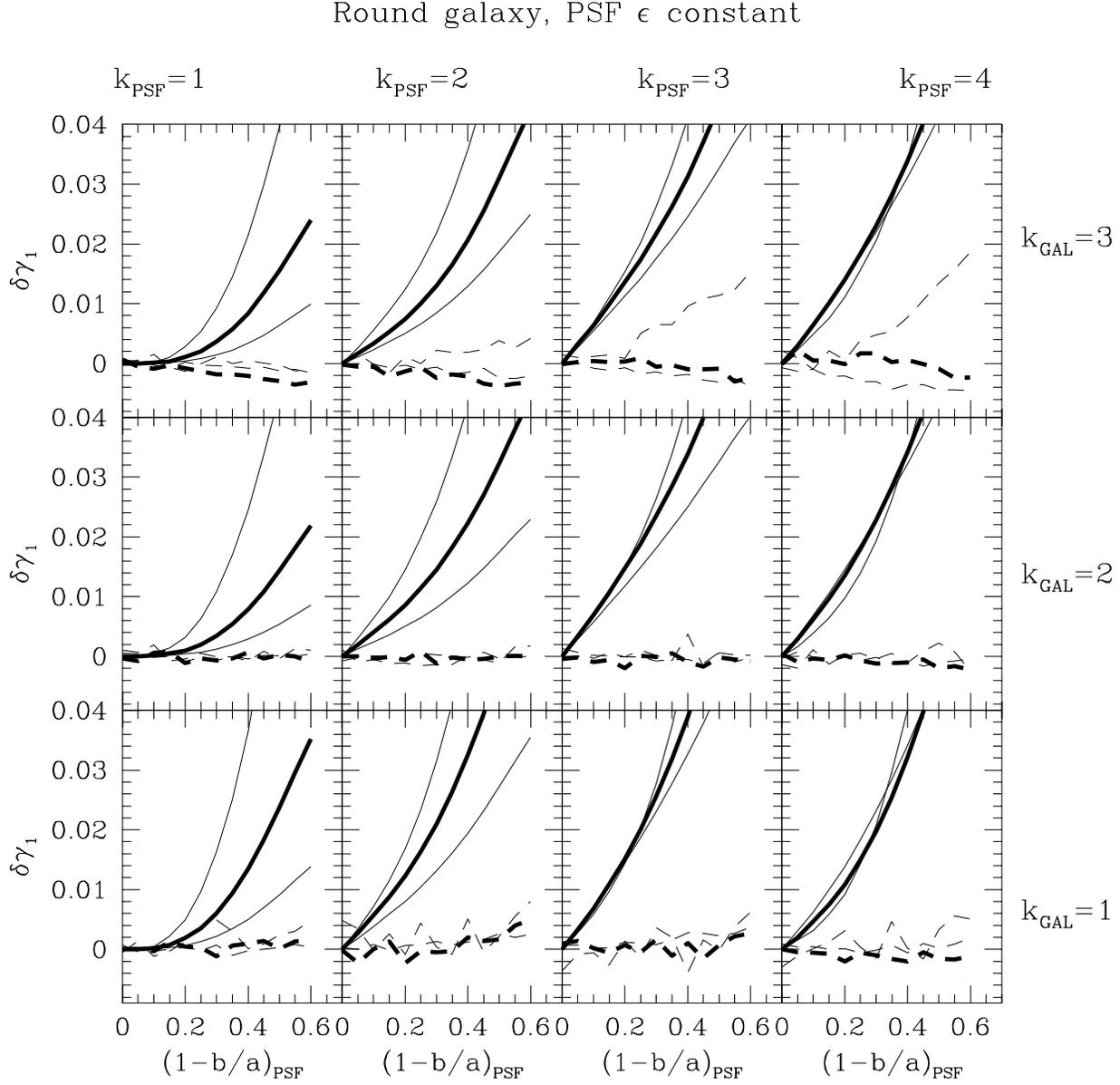}}
\caption{The result of correcting simulated unsheared images for PSF
anisotropy, following the KSB method (solid lines) and the method
presented here (dashed lines). $\delta\gamma_1$ is in each case the
shear that is deduced after the PSF correction, and should be zero for
a perfect analysis.  The $k$'s are luminosity profile shape parameters
for galaxy and PSF, and are explained in the text. In each panel the
heavy line represents the case where the galaxy image is intrinsically
of the same radius as the PSF ($r_g=r_p$ in eqs. \ref{eq:gal} and
\ref{eq:psf}), the lighter lines those where the galaxy is 0.5 (upper)
and 1.5 (lower) times this size. The PSF ellipticity is constant with
radius in these simulations. While for the $k_{\rm PSF}=1$ case
(Gaussian PSF) the KSB method leaves a residual which is third-order
in PSF ellipticity, other PSF luminosity profiles give rise to
first-order residuals.
The residuals of the new method in the upper panels disappear if more
radial components are used in the fit for $\overline{g}$, highlighting
that the dominant source of error in this method is the extent to
which the radial profile is modeled correctly.
}
\label{fig:rnda}
\end{figure*}
\begin{figure*}
\resizebox{\hsize}{!}{\includegraphics{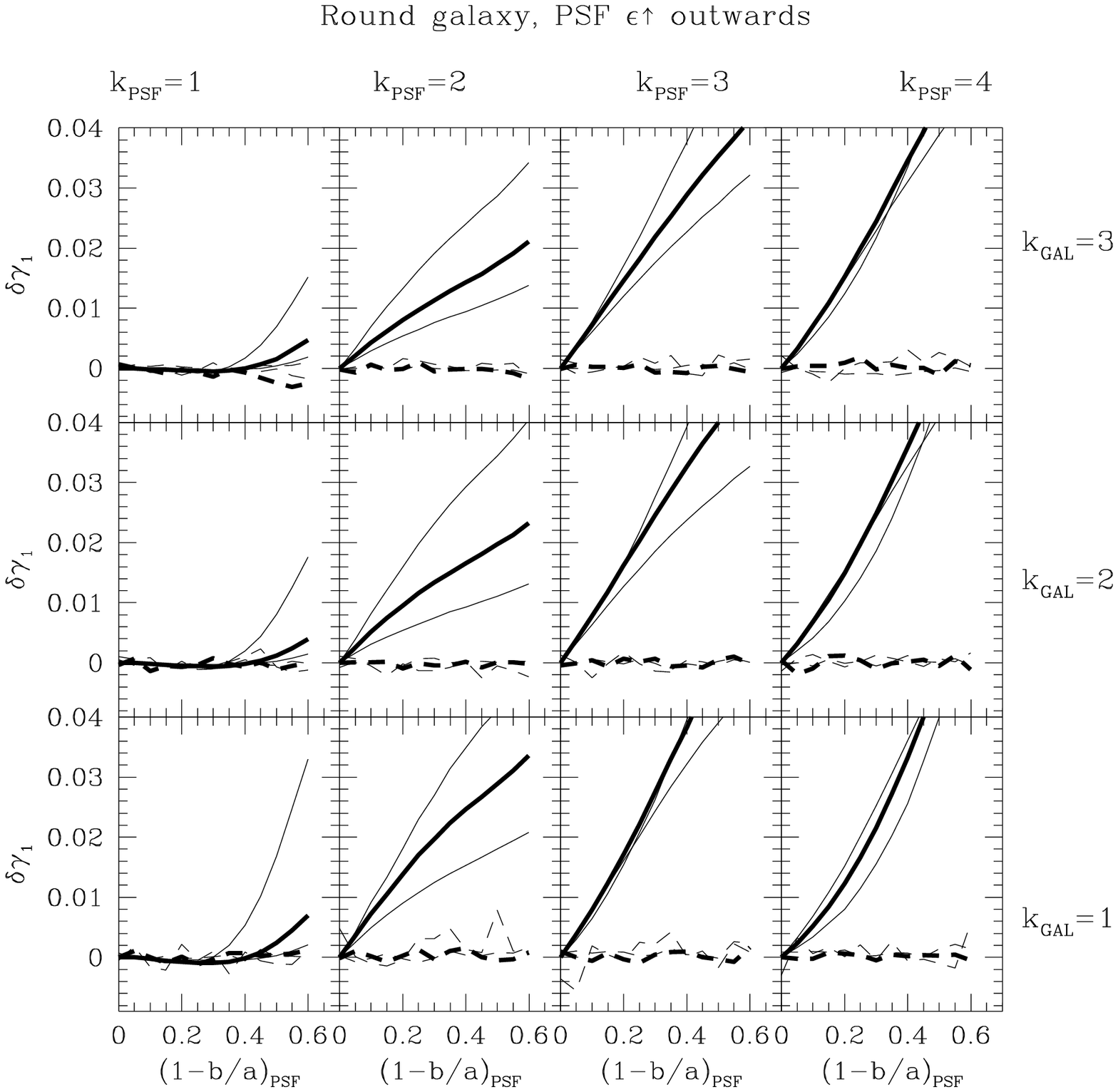}}
\caption{As figure \ref{fig:rnda}, but only the outer component of the
PSF is elliptical.}
\label{fig:rndb}
\end{figure*}
\begin{figure*}
\resizebox{\hsize}{!}{\includegraphics{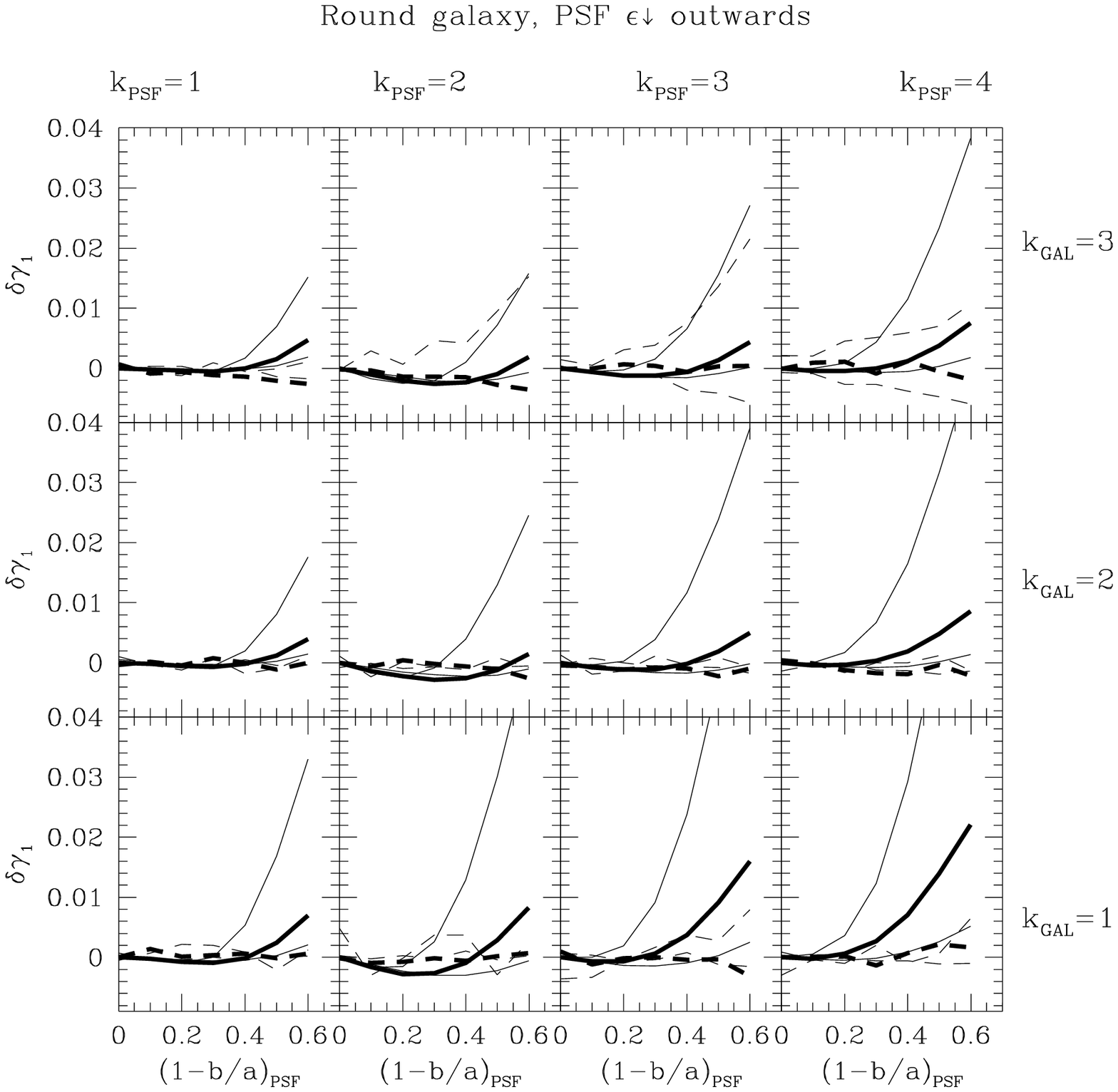}}
\caption{As figure \ref{fig:rnda}, but only the inner component of the
PSF is elliptical.}
\label{fig:rndc}
\end{figure*}

The advantage of the multiple-Gaussian formulation is that the PSF
convolution can be done analytically. We thus constructed a large
number of PSF-smeared galaxy images, calculated the polarizations and
polarizabilities following HFKS, subtracted the PSF anisotropy correction 
\[\delta e=P^{{\rm sm}\, g} (P^{\rm sm\,\star})^{-1}e^\star\]
 to the galaxy polarization, and divided the result by Luppino and
Kaiser (1997)'s pre-seeing shear polarizability $P^\gamma$.  We then
compared these results with the results of our implementation of the
new fitting algorithm of Figure~\ref{fig:algo}.

The results of the simulations are
presented in figures \ref{fig:rnda}, \ref{fig:rndb} and
\ref{fig:rndc}. 
They show that the KSB method can suffer from systematic
residuals around the 0.01 shear level once the PSF ellipticity
exceeds 0.2 or so, whereas this is not so for the new method developed
here. The systematic effects are strongest for small galaxies, for PSF
profiles with long tails, and for radially increasing PSF
ellipticity. The effect is clearly driven by the PSF shape,
not by the galaxy brightness profile.

Notice that in the constant-ellipticity case (Figure~\ref{fig:rnda}),
with a Gaussian PSF ($k_{\rm psf}=1$) the residuals left by the KSB
method are high order in PSF ellipticity, but that for non-Gaussian
PSF's a low-order residual dominates. (We have verified this result
analytically using symbolic mathematics.) This is a consequence of the
fact that only the single elliptical Gaussian PSF can be written as a
convolution of a compact anisotropic function with a round extended
one, as assumed in the KSB derivation. It is clearly important to test
algorithms not only for single-Gaussian PSF's!

\subsubsection{A WFPC-2 PSF}

\begin{figure}
\resizebox{\hsize}{!}{\includegraphics{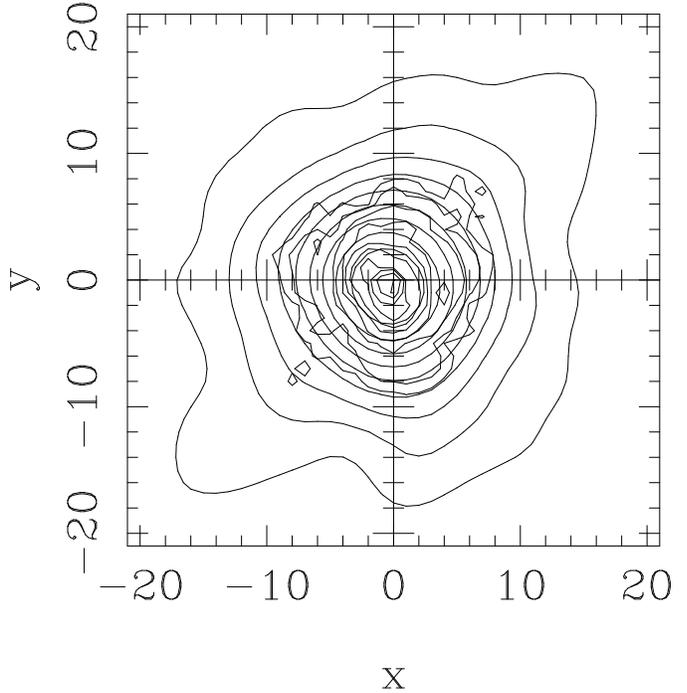}}
\caption{PSF and simulated galaxy image for an observation in the
corner of one of the WFPC2 CCD's. Axis units are 0.05 arcsec.}
\label{fig:wfpc}
\end{figure}

In order to test whether our results are specific to the
double-Gaussian formulation of the PSF, a test was also performed with
a model PSF for the WFPC-2 camera on the Hubble Space Telescope. The
model was generated with the Tiny TIM software package, provided
on-line at STScI by J.~Krist. An oversampled PSF was calculated for a
position near the corner of CCD\#4, and convolved with a Gaussian
circular galaxy of FWHM 0.25arcsec. This `galaxy' and the PSF
(Figure~\ref{fig:wfpc}) were then binned to a resolution of half a
WFPC-2 pixel to avoid under-resolving the PSF, and analyzed as
above. The results are summarized in table~\ref{tab:wfpcsim}, and
confirm the results obtained from the large number of double-Gaussian
simulations described earlier.

\begin{table}
\caption[Simulation with a WFPC-2 PSF]
{Results of a simulation based on a WFPC-2 PSF calculated
using the TinyTim software. In agreement with earlier results (HFKS),
the KSB technique appears to over-correct for the anisotropic WFPC-2
PSF images slightly.}
\label{tab:wfpcsim}
\begin{tabular}{ll}
Weight function radius: &1.3 WFPC-2 pixels\\
PSF polarization:&$(-0.078,0.024)$\\
Uncorrected galaxy $e$:&$(-0.020,-0.006)$\\
PSF-corrected galaxy $\gamma$:&$(-0.000,-0.019)$ (KSB)\\
&$(-0.000,-0.001)$ (new method)\\
\end{tabular}
\end{table}

\subsection{Noise properties}

\subsubsection{Analytic estimate}
\label{sec:noise-an}

The error on the estimated shear due to photon noise can be estimated as
follows. Let the 1-$\sigma$ error on each pixel of $\overline g$ be
$\sigma$ (for simplicity we take this to be the same on every pixel,
appropriate for background-limited work). Then the fit involves
finding the minimum of
\begin{equation}
\chi^2=\sum_k\left[\overline{g}_k-(P\otimes f({\mathbf
x}\cdot{\mathbf\Gamma}^2\cdot{\mathbf x}))_k\right]^2/\sigma^2
\end{equation}
where $P(x,y)$ is the PSF, $f(r^2)$ is the intrinsic radial profile of
the average galaxy, $\otimes$ denotes convolution, ${\mathbf x}_k$ is
the position of the $k$th pixel, and ${\mathbf\Gamma}$ is the
distortion matrix of equation~\ref{eq:dist}.  If the fit parameters
$\gamma_i$ are uncorrelated with the radial profile, their inverse
variances are given by $\frac12\partial^2\chi^2/\partial \gamma_i^2$. For
example, at the best fit
\begin{eqnarray}
\frac12{\partial^2\chi^2\over\partial \gamma_1^2}
=&\displaystyle
\sum_k\left({\partial\over\partial \gamma_1}
(P\otimes f({\mathbf x}\cdot{\mathbf\Gamma}^2\cdot{\mathbf
x}))_k\right)^2/\sigma^2,\cr
=&\displaystyle
\sum_k\left(P\otimes [4f'(r^2)(y^2-x^2)]\right)^2/\sigma^2.
\label{eq:noise-an}
\end{eqnarray}
The right-hand side of equation~\ref{eq:noise-an} can be estimated
assuming that the PSF and observed average galaxy are Gaussians with
dispersions $r_{\rm PSF}$ and $r_{\rm GAL}$ pixels of integral 1 and
$F$, respectively.  Then the 1-$\sigma$ error on $\gamma_1$ evaluates to
\begin{eqnarray}
\sigma(\gamma_1)=&
\displaystyle
\left(\frac12{\partial^2\chi^2\over\partial
\gamma_1^2}\right)^{-1/2}
={2\pi^{1/2}r_{\rm GAL}^3\sigma \over (r_{\rm GAL}^2-r_{\rm PSF}^2)F}\cr
=&
\displaystyle
{r_{\rm GAL}^2\over (r_{\rm GAL}^2-r_{\rm PSF}^2)}{\delta F\over F},
\label{eq:noise-an2}
\end{eqnarray}
where we have used the results that the PSF-fitting error on $F$ for a
Gaussian source is $\delta F=2\pi^{1/2}r_{\rm GAL}\sigma$. (The error
on $\gamma_2$ is the same.) We have verified this formula by means of
simulations, similar to those described below. Equation
\ref{eq:noise-an2} shows the expected increase in noise for small
objects, as well as the limit, even for large objects, of
\begin{equation}
\sigma(\gamma_i)\la{\delta F\over F}.
\end{equation}


\subsubsection{Simulations}

We have checked the sensitivity to noise in the images by Monte Carlo
simulation. Many realizations of random Gaussian noise superimposed on
a PSF-smeared, intrinsically round galaxy image were analyzed
with both algorithms, and the distributions of the resulting
$(\gamma_1,\gamma_2)$ estimates  
compared. Selected results are shown in Table~\ref{tab:noise}.
Interestingly, the dispersions in the shears derived with both methods
are very similar over a range of galaxy sizes. As we have
already seen, the small bias in the results from KSB is present in the
simulations with non-circular PSF's, but not in the method advocated in
this paper.

A possible way to avoid the systematic residuals of the KSB method is
to increase the radius of the weight function $W$ in
eq.~\ref{eq:wtfn}, since the problems arise from the imperfect way in
which the polarizabilities represent the effect of $W$. However, the
primary function of $W$ is to control the noise in the
images. Doubling the Gaussian radius of $W$ does in fact improve the
anisotropy correction in the mean, but at the cost of almost doubling
the noise on the result (see Table~\ref{tab:noise}). 

A by-product of our algorithm is an estimate of the intrinsic radial
profile of $\overline{g}$. In practice, this estimate appears to be
rather sensitive to the noise, especially for small images---not
surprising given that this is effectively a deconvolution, albeit a
constrained one.  Nevertheless, it may be possible to use the
information in the best-fit radial profile in several ways. If a
suitable prior for the intrinsic radial profile of the average galaxy
selected can be formulated (e.g., by combining results over a wide
field, or from deeper, higher resolution images), this information
might help to refine the best-fit shear solution
further. Alternatively, the width of $\overline{g}$ might be used to
attempt to derive the lensing convergence $\kappa$ directly, since in
principle it is a direct measure of the magnification of faint
galaxies.  This possibility is yet to be explored in detail, but is
likely to be difficult in practice.

\begin{table*}
\caption[Noise Simulations] {Results from representative noise
simulations of the KSB method and the one presented in this paper. In
each case, 100 noise realizations (noise per pixel of 0.001, with
$r_{\rm psf}=2$pixels, $k_{\rm gal}=2$ and total flux 4) were analyzed
with the standard KSB method, with the KSB method using a weight
function double the radius of the best-fit Gaussian, and with the new
method described in this paper. The first six simulations were of
cases without PSF anisotropy, and in the last six the PSF has a
constant axis ratio of 0.7. In all cases, the dispersions of the
standard KSB method and the new one are very similar, but note the
imperfect correction from the KSB method. The simulations with a
larger weight function show that the PSF anisotropy is corrected
better, but at the cost of increased noise.}
\label{tab:noise}
\begin{tabular}{ll@{$\qquad$}ll@{$\qquad$}ll@{$\qquad$}l}
\multicolumn{4}{c}{KSB$\quad$}&
\multicolumn{2}{c}{This paper$\quad$}\\
\multicolumn{2}{c}{Standard $W$$\quad$}&\multicolumn{2}{c}{Wider $W$$\quad$}&
\multicolumn{2}{c}{(4 radial cpts.)$\quad$}\\
Mean & \multicolumn{1}{c}{$\sigma$} &
Mean & \multicolumn{1}{c}{$\sigma$} &
Mean & \multicolumn{1}{c}{$\sigma$} &Comments\\
  0.0010 &  0.0084 &   0.0016 &  0.0142 & 0.0010  & 0.0079 & 
Round Gaussian PSF, $r_{\rm gal}=0.5r_{\rm PSF}$\\
  0.0008 &  0.0067 &   0.0009 &  0.0105 & 0.0008  & 0.0061 & 
Round Gaussian PSF, $r_{\rm gal}=r_{\rm PSF}$\\
  0.0008 &  0.0072 &  0.0007 &  0.0113 &  0.0009  & 0.0066 & 
Round Gaussian PSF, $r_{\rm gal}=1.5r_{\rm PSF}$\\
  0.0019 &  0.0149 &  0.0031 &  0.0310 &  0.0023  & 0.0150 & 
Round $k=3$ PSF, $r_{\rm gal}=0.5r_{\rm PSF}$\\
  0.0012 &  0.0104 &  0.0014 &  0.0206 &  0.0017  & 0.0110 & 
Round $k=3$ PSF, $r_{\rm gal}=r_{\rm PSF}$\\
  0.0012 &  0.0105 &  0.0012 &  0.0186 &  0.0014  & 0.0107 & 
Round $k=3$ PSF, $r_{\rm gal}=1.5r_{\rm PSF}$\\
  0.0116 &  0.0056 &  0.0045 &  0.0097 &  0.0011  & 0.0061 & 
$\epsilon=0.3$ Gaussian PSF, $r_{\rm gal}=0.5r_{\rm PSF}$\\
  0.0041 &  0.0054 &   0.0017 &  0.0087 & 0.0009  & 0.0054 & 
$\epsilon=0.3$ Gaussian PSF, $r_{\rm gal}=r_{\rm PSF}$\\
  0.0020 &  0.0064 &  0.0009 &  0.0102 &  0.0008  & 0.0060 & 
$\epsilon=0.3$ Gaussian PSF, $r_{\rm gal}=1.5r_{\rm PSF}$\\
  0.0277 &  0.0101 &  0.0222 &  0.0214 &  0.0023  & 0.0116 & 
$\epsilon=0.3$ $k=3$ PSF, $r_{\rm gal}=0.5r_{\rm PSF}$\\
  0.0244 &  0.0085 &  0.0112 &  0.0162 &  0.0007  & 0.0091 & 
$\epsilon=0.3$ $k=3$ PSF, $r_{\rm gal}=r_{\rm PSF}$\\
  0.0191 &  0.0091 &  0.0061 &  0.0157 &  0.0008  & 0.0094 & 
$\epsilon=0.3$ $k=3$ PSF, $r_{\rm gal}=1.5r_{\rm PSF}$\\
\end{tabular}
\end{table*}

\subsubsection{The effect of centroiding errors}
The centroid of an image can be determined in different ways, each of
them susceptible to errors due to photon noise. The effect of
centroiding errors on the summed galaxy image will be a convolution
with the distribution of centroid errors. Thus, the PSF needs to be
convolved with this distribution before analysis of $\overline g$, so
that the effect of the centroiding error can be compensated.

\section{Galaxy-by-galaxy application}

The method as described so far involves analysing the average galaxy
$\overline g$. Very accurate shear measurements require $\overline g$
to be the average of a large number of galaxies ($\sim1000$ for a
1-$\sigma$ shear accuracy of 0.01), otherwise intrinsic ellipticity
scatter will dominate. However, constructing such a $\overline g$ is
only possible if the shear and the PSF are constant over a large part
of the image. Often this is not the case.

To cope with this limitation, we have therefore experimented with the
algorithm in `galaxy-by-galaxy' mode, where the algorithm is applied
to individual galaxies and the resulting shear estimates averaged.
Mathematically this approach is not perfect, because it involves
fitting a constant-ellipticity model to individual galaxies even
though this is not necessarily appropriate. Nevertheless it turns out
to work better than might have been expected, and better than existing
methods.

We tested this approach on various model galaxies, of differing axis
ratios. To simulate typical galaxies, we include a round, central
`bulge' component, and an outer `disk' of axis ratio between 0.1 and
1. (Simulations with different bulge axis ratios yielded the same
results.) These were placed at all orientations, smeared with an
elliptical PSF, and analysed with the algorithm described above. The
best-fit $(\gamma_1,\gamma_2)$ values thus derived for each galaxy are
then averaged to give an estimate of the shear.

\begin{figure*}
\resizebox{\hsize}{!}{\includegraphics{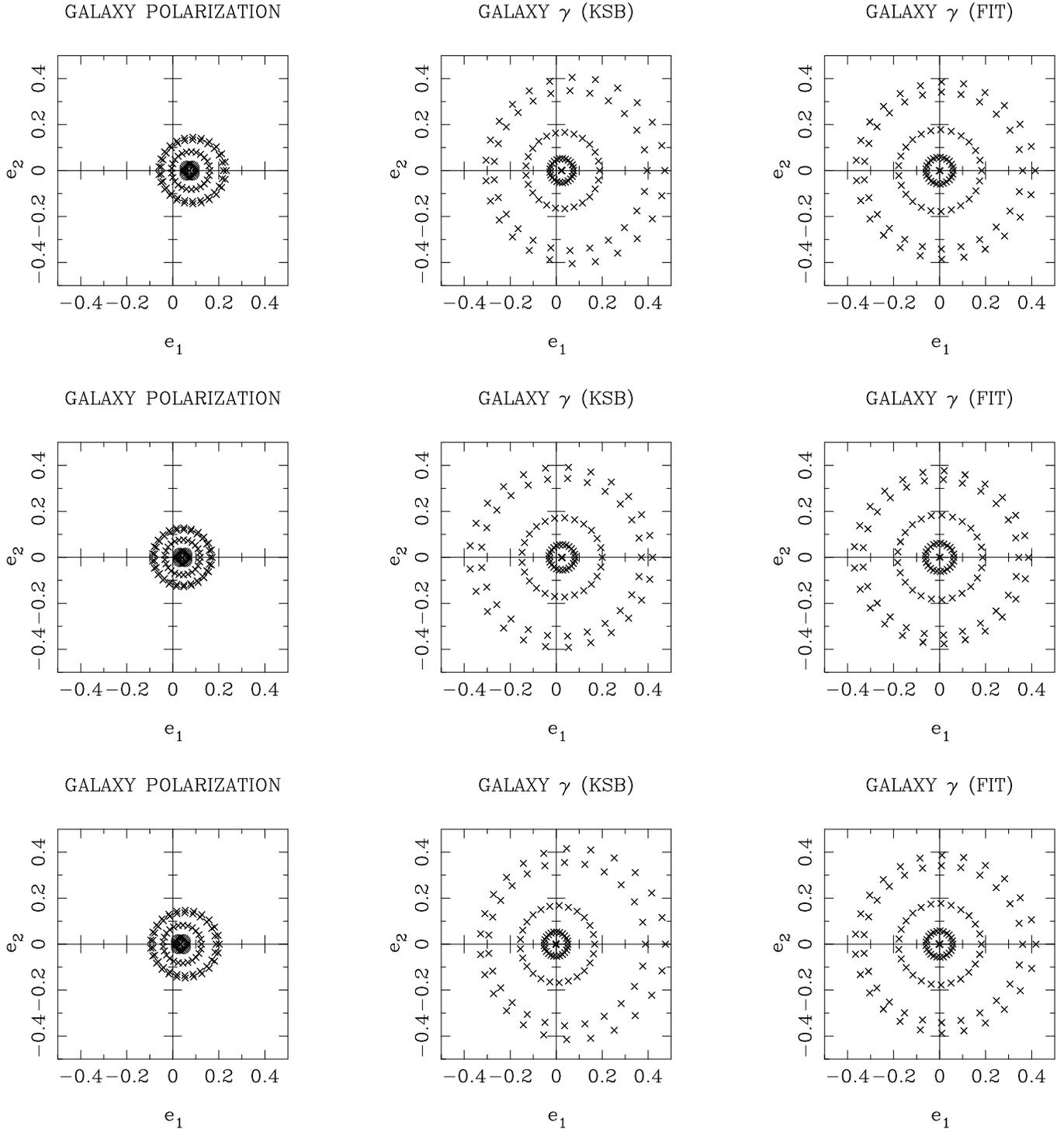}}
\caption{The derived shear values from the algorithm applied to
individual elongated `disk+bulge' galaxy images, after smearing with
an elongated PSF. Each ring represents a galaxy of different shape,
seen at many orientations. Left: the raw $(e_1,e_2)$ polarizations measured
with the standard KSB method, without correction for the PSF. Centre:
the result of applying the KSB prescription for PSF anisotropy and
circularization correction. The small bias seen before remains. Right:
the result of the new algorithm on the same galaxies. In the latter
case, the correct
average shear estimate (zero) is recovered even though individual
galaxies are not correctly described as intrinsically
constant-ellipticity sources. The three rows refer to the three kinds
of PSF ellipticity profile considered in 
figures \ref{fig:rnda}--\ref{fig:rndc}: constant with radius (top), 
outward-increasing (middle), and outward-decreasing (bottom).}
\label{fig:galbygal}
\end{figure*}

As may be seen in Figure~\ref{fig:galbygal}, the algorithm performs
very well, essentially correcting all PSF anisotropy signal in the
measured shear. By comparison the slightly biased answer returned by
the KSB algorithm is apparant as before. Residual systematics of the
new method are at the level of a few tenths of a percent.

\section{Summary}

In this paper we have studied possible systematic errors arising from
the correction for anisotropic point-spread functions in weak lensing
analyses based on the well-known Kaiser et al.\ (1995) method. While
such effects are small, generally below a few percent in the deduced
gravitational shear components $(\gamma_1,\gamma_2)$, they are at a
level that is important for studies such as galaxy-galaxy
lensing, lensing by large-scale structure or cluster lensing at large
radii. A range of simulations shows that modelling the PSF as
a convolution of a compact anisotropic function with a more extended,
circular function, which underlies the KSB
formulation, is not sufficiently general to describe many PSFs, and
leads to these systematic residuals.

We have presented a new algorithm with which to carry out the
PSF-correction in a single fitting step, and show with simulated
images that the low-level residuals left in the KSB analysis can thus
be avoided. The whole image is used in the fitting, so that not just
the lowest moments are used to characterize the image shapes. We have
also shown that the noise properties of this algorithm compare well
with those of KSB. While mathematically the algorithm requires an
intrinsically circular source, as may be constructed by stacking many
observed galaxy images, in practice nearly unbiased results can also
be obtained when the algorithm is used to correct individual galaxy
shapes for the PSF. This galaxy-by-galaxy application of the algorithm
allows observations with spatially varying PSF and/or shear fields to
be handled. 

Weak lensing is a unique technique with which to study gravitational
potentials at large radii in galaxy clusters, galaxy halos and in the
field. The present method holds the promise of allowing a little more
information to be extracted from the large volumes of data that will
be gathered with the coming generation of wide-field imagers.

\begin{acknowledgements}
I would like to thank Peter Schneider, Marijn Franx, Henk Hoekstra and
the referee for critical readings of the manuscript and for suggesting
several improvements.
\end{acknowledgements}


\begin{thebibliography}{}
\bibitem[1993]{q2345}
Bonnet, H., Fort, B., Kneib, J.-P., Mellier, Y., Soucail, G., 1993,
A\&A 280, L7
\bibitem[1995]{bm}
Bonnet, H., Mellier, Y. 1995, A\&A 303, 331
\bibitem[1996]{brainerd}
Brainerd, T., Blandford, R.D., Smail, I., 1996, ApJ 466, 623
\bibitem[1998]{clowe} Clowe, D., Luppino, G. A., Kaiser, N., 
Henry, J. P., \& Gioia, I. M., 1998, ApJ 497, L61 
\bibitem[1994]{fahlman}
Fahlman, G., Kaiser, N., Squires, G., Woods, D., 1994, ApJ 437, 56
\bibitem[1997]{q0957} Fischer, P., Bernstein, G., Rhee, G., Tyson,
J. A., 1997, AJ 113, 521
\bibitem[1997]{ft97} Fischer, P., Tyson, J.  A., 1997, AJ 114, 14
\bibitem[1998]{hoekstra}
Hoekstra, H., Franx, M., Kuijken, K., Squires, G., 1998, ApJ, in press
\bibitem[1997]{jain97} Jain, B., Seljak, U., 1997, ApJ 484, 560 
\bibitem[1998]{kaiser98} Kaiser, N., 1998, ApJ 498, 26 
\bibitem[1999]{k99} Kaiser, N., 1999, preprint.
\bibitem[1995]{ksb}
Kaiser, N., Squires, G., Broadhurst, T., 1995, ApJ 449, 460
\bibitem[1998]{lb} Lombardi, M. \& Bertin, G., 1998, A\&A, 330, 791
\bibitem[1997]{lk97} Luppino, G. A., Kaiser, N., 1997, ApJ 475, 20 
\bibitem[1998]{schneider98} Schneider, P., Van Waerbeke, L., 
Mellier, Y., Jain, B., Seitz, S., Fort, B., 1998, A\&A 333, 767 
\bibitem[1997]{a2163}
Squires, G., et al., 1997, ApJ 482, 648
\bibitem[1990]{tvw}
Tyson, J., Valdes, F., Wenk, R., 1990, ApJ 349, L1
\bibitem[1997]{waerbeke}
Van Waerbeke, L., Mellier, Y., Schneider, P., Fort, B., Mathez, G., 1997,
A\&A 317, 303


\end{thebibliography}
\end{document}